\documentclass[authorversion,nonacm, sigconf, screen]{acmart}
\AtBeginDocument{%
  \providecommand\BibTeX{{%
    \normalfont B\kern-0.5em{\scshape i\kern-0.25em b}\kern-0.8em\TeX}}}

\setcopyright{none}

\usepackage{adjustbox}
\usepackage{booktabs}
\usepackage{amsmath} % Essential for math environments
\usepackage{mathtools}
\mathtoolsset{showonlyrefs}
\usepackage{array}
\DeclareMathOperator*{\argmax}{arg max}

\usepackage{ifthen}
\newboolean{anonym}
\setboolean{anonym}{false}

\begin{document}

\title{FinDKG: Dynamic Knowledge Graphs with Large Language Models for Detecting Global Trends in Financial Markets}

\author{Xiaohui Victor Li}
\affiliation{%
  \institution{Imperial College London}
  \country{London, United Kingdom}
}
\email{xiaohui.li21@alumni.imperial.ac.uk}
\authornote{This work was completed as part of XVL's MSc project at Imperial College London.}

\author{Francesco Sanna Passino}
\affiliation{%
  \institution{Imperial College London}
  \country{London, United Kingdom}
}
\email{f.sannapassino@imperial.ac.uk}

\begin{abstract}
Dynamic knowledge graphs (DKGs) are popular structures to express different types of connections between objects over time. They can also serve as an efficient mathematical tool to represent information extracted from complex unstructured data sources, such as text or images. Within financial applications, DKGs could be used to detect trends for strategic thematic investing, based on information obtained from financial news articles.
In this work, we explore the properties of large language models (LLMs) as dynamic knowledge graph generators, proposing a novel open-source fine-tuned LLM for this purpose, called the Integrated Contextual Knowledge Graph Generator (ICKG). We use ICKG to produce a novel open-source DKG from a corpus of financial news articles, called FinDKG, and we propose an attention-based GNN architecture for analysing it, called KGTransformer.  
We test the performance of the proposed model on benchmark datasets and FinDKG, demonstrating superior performance on link prediction tasks. Additionally, we evaluate the performance of the KGTransformer on FinDKG for thematic investing, showing it can outperform existing thematic ETFs.
\end{abstract}


\maketitle

%%% > Introduction section
\section{Introduction}

A knowledge graph (KG) is a data structure that encodes information consisting in entities and different types of relations between them. Formally, a KG can be represented as $\mathcal{G} = \{ \mathcal E, \mathcal R, \mathcal F \}$, where $\mathcal E$ and $\mathcal R$ denote the sets of entities and relations respectively, and $\mathcal F\subseteq \mathcal E\times\mathcal R\times\mathcal E$ represents a set of facts, consisting in relations of different types between entities. %The set $E$ contains discrete units of knowledge or entities. The set $R$ consists of various heterogeneous relations representing connections between entities. Lastly, $F$ comprises events, corresponding to factual information typically represented as a triplet, linking entities and relations.
The triplet $(s, r, o) \in \mathcal F$ is the fundamental building block of a KG, where $s \in \mathcal E$ represents the source entity, $r \in \mathcal R$ the relation, and $o \in \mathcal E$ the object entity. For instance, the triplet \textit{(OpenAI, Invent, ChatGPT)} shows how entities and relations combine to form a fact, with \textit{OpenAI} and \textit{ChatGPT} as entities and \textit{Invent} as the relation. %When these fact triplets are aggregated into a comprehensive KG, they enable a collective understanding and interpretation of globally structured knowledge.

Temporal or dynamic knowledge graphs (DKGs)
extend static KGs by incorporating temporal dynamics. Each fact in a DKG is associated with a timestamp $t \in \mathbb{R}_+$, allowing the model to capture the temporal evolution of events. %, as shown in Figure~\ref{fig:DKG}. 
%Let $N(t)$ denote the cumulative number of facts up to time $t\in\mathbb R_+$, with each fact assigned an %ordering 
%index $i\in\mathbb N$ such that $t_i\leq t_j$ for $i<j,\ i,j\in\mathbb N$. 
Therefore, events occur in quadruples $(s_i,r_i,o_i,t_i)\in\mathcal{E}\times\mathcal{R}\times\mathcal{E}\times\mathbb R_+$, where $t_i$ is the event time, such that $t_i\leq t_j$ for $i<j,\ i,j\in\mathbb N$. 
Then, the DKG $\mathcal{G}_t=(\mathcal E,\mathcal R,\mathcal F_t)$ at time $t$ can be expressed via a time-varying set of facts $\mathcal F_t$ defined as
\begin{equation}
\mathcal F_t = \{(s_i, r_i, o_i, t_i) : s_i,o_i\in \mathcal E,\ r_i\in \mathcal R,\ t_i< t\}. \label{eq:Ft}
\end{equation}
The task of estimating a model for $\mathcal G_t$ from observed data is called \textit{dynamic knowledge graph learning}. %The objective of dynamic knowledge graph learning is to approximate the underlying distribution of temporal graphs  \( p(\mathcal{G}_{t}) \) at a given timestamp as \( t \). 
This typically involves data-driven training of graph neural networks, designed to model both the structure and the temporal dynamics of the KGs over time.

In real-world applications such as finance, entities and relations can be further grouped into \textit{categories}, often called \textit{meta-entities}. For example, consider the relation between the entity \textit{Jeff Bezos} which is of type \textit{Person}, and the entity \textit{Amazon}, which is of type \textit{Company}. The relation between them is \textit{Founder Of}, which could be considered to have the type \textit{Business action}. In this work, inspired by heterogeneous graph transformers \citep[HGT,][]{hu2020heterogeneousgraphtransformer}, we discuss a way to introduce the additional meta-entity information within a dynamic knowledge graph learning procedure based on graph attention networks \citep[GAT,][]{velivckovic2017graph} and EvoKG \citep{park2022evokg}. This results in the \textit{Knowledge Graph Transformer} (\textit{KGTransformer}), an attention-based graph neural network (GNN) designed to create dynamic lower-dimensional representations of entities and relations. %, known as embeddings.

In addition to DKGs, Large Language Models (LLMs) have also %emerged as transformative tools 
been gaining popularity recently within the financial sector, demonstrating potential in enhancing various financial tasks through advanced natural language processing (NLP) capabilities \cite{nie2024survey}. Popular models such as BERT, the GPT series, and financial-specific variants such as FinBERT \cite{araci2019finbert} and FinGPT \cite{yang2023fingptopensourcefinanciallarge} leverage LLMs to improve the state-of-the-art in tasks such as financial sentiment analysis.

The application of LLMs to dynamic knowledge graphs has been so far limited in the literature. Therefore, one of the main contributions of this work is to also propose a pipeline for generative knowledge graph construction (KGC) via Large Language Models (LLMs), resulting in the  \textit{Integrated Contextual Knowledge Graph Generator (ICKG)} large language model.  %Prompt engineering is a technique that manipulates the input query or \textit{``prompt''} to better guide the model's response. The process of generative Knowledge Graph Construction (KGC) refers to employing 
In particular, we develop a fine-tuned LLM to systematically extract entities and relationships from textual data via %effective prompts
engineered input queries or \textit{``prompts''}, subsequently assembling them into event quadruples of the same form as \eqref{eq:Ft}.
We use the proposed ICKG LLM to generate an open-sourced financial knowledge graph dataset, called FinDKG. 

In summary, our contributions in this work are threefold:
\begin{enumerate}
\item We propose KGTransformer, an attention-based GNN architecture for dynamic knowledge graph learning that includes information about meta-entities (\textit{cf.} Section~\ref{sec:learning}),
combining existing work on GATs \citep{velivckovic2017graph}, HGTs \citep{hu2020heterogeneousgraphtransformer} and EvoKG \citep{park2022evokg}. %, a model for DKGs.
We demonstrate substantial improvements in link prediction metrics (\textit{cf.} Section~\ref{sec:lp}) on real-world DKGs.
\item We develop an open-source LLM for dynamic knowledge graph generation for finance called \textit{Integrated Contextual Knowledge Graph Generator} (ICKG, \textit{cf.} Section~\ref{sec:ickg}).
\item We utilise ICKG to create an open-source dynamic knowledge graph based on financial news articles, called \textit{FinDKG} (\textit{cf.} Section~\ref{sec:findkg}). FinDKG is used for thematic investing upon capitalizing on the AI trend, %demonstrating significant improvements over 
improving upon other AI-themed portfolios (\textit{cf.} Section~\ref{sec:findkg_investing}).
\end{enumerate}

The remainder of this work is organised as follows: Section~\ref{sec:literature} discusses related literature. Next, Section~\ref{sec:ickg} and~\ref{sec:learning} discuss the main contributions of our work: ICKG and KGTransformer. Finally, Section~\ref{sec:experiments} discusses applications on real-world DKGs. %, followed by a conclusion and discussion in Section~\ref{sec:conclusion}.  

\section{Related literature}
\label{sec:literature}

\paragraph{Graph representation learning}
Graph representation learning via graph neural networks (GNNs) is a fast-growing branch of deep learning, focused on extracting lower-dimensional latent space representations of graphs, to improve performance in downstream applications \cite{chen2020graph}. These methods have demonstrated significant capabilities in tasks such as node classification, edge prediction, and graph classification \cite{kipf2016semi, xu2018powerful, khoshraftar2024survey}. When applied to knowledge graphs, representation learning is aimed at deriving low-dimensional vector representations of entities and relations \cite{ji2021survey}, called embeddings. %These embeddings are then used for %downstream 
Within the context of KGs, embeddings are then used for
tasks such as information retrieval \cite{reinanda2020knowledge}, question answering \cite{bordes2015large}, and recommendations \cite{wang2018ripplenet,wang2019knowledge}. 
Recent advancements in temporal knowledge graph learning have also integrated temporal information %making this approach highly effective for modeling real-world datasets in time-aware applications 
\citep{cai2024survey}.

\paragraph{Financial knowledge graphs}
Financial systems are often characterised by intricate and dynamically evolving relationships \cite{acemoglu2016networks}, which can be represented as DKGs for applications such as fraud transaction identification \cite{weber2019anti}, stock return prediction \cite{feng2019temporal}, stock linkage discovery \cite{chung2023modeling}, and network-based portfolio construction \cite{turner2023graph}. However, the heterogeneous and dynamic nature of financial networks poses challenges for existing static GNN models, and the study of dynamic extensions of these models within a financial context remains relatively underdeveloped, despite advancements in financial natural language processing \cite{gentzkow2019text}.
Early industry applications of financial KGs were based on static knowledge graph models \cite{fu2018stochastic, cheng2020knowledge}.
Also, \cite{yang2020knowledge} highlighted the potential of KGs in finance by developing a static macroeconomics knowledge graph for selecting variables in economic forecasting. Their KG-based methods improved forecasting accuracy.
In this work, we propose an architecture which incorporates \textit{meta-entities} within DKGs, and demonstrate its performance on finance-related tasks. % such as thematic investing. 

\paragraph{LLMs in finance}

LLMs have been applied to a wide array of financial tasks. %One of the application focus is sentiment analysis, these models excel in interpreting and quantifying market sentiment from diverse textual sources such as financial news, social media, and corporate disclosures. 
For example, \cite{araci2019finbert} and \cite{yang2023fingptopensourcefinanciallarge} demonstrate the effectiveness of LLMs in extracting sentiment from financial news, social media, and corporate disclosures. %, leading to improved market prediction accuracy.
%In financial time series analysis, LLMs have shown notable advancements. Their deep learning architectures enable the capture of complex temporal dependencies and patterns within financial datasets. %\cite{chen2022expected} fine-tuned LLMs to time series forecasting, achieving state-of-the-art results in stock price prediction; Additionally, 
\cite{lopez2023can} demonstrates good performance of GPT-4 %\cite{openai2023gpt} 
in predicting stock market returns based on financial news headlines, claimed to be superior to sentiment analysis.
Despite these advancements, challenges such as interpretability and computational costs with closed-sourced LLMs remain. %Addressing these issues is essential to ensure the responsible and effective deployment of LLMs in finance. 
\cite{inserte2024large} emphasises the need for improved interpretability in LLMs to promote transparency for financial applications. 
Moreover, while existing commercial LLMs such as GPT-4 offer substantial capabilities, their closed-source architecture imposes constraints on their usage. 
%GPT-4 demonstrates a higher quality response in adhering to instruction prompts for KG triplet extraction. Nevertheless, the financial overhead and computational latency associated with GPT-4's API constitute realistic obstacles.
Open-source models such as Meta's LLaMA \cite{LLaMa} and Mistral AI's LLM \cite{Jiang23} %democratize access by offering 
offer more efficient alternatives, albeit often less precise.

\section{The Integrated Contextual Knowledge Graph Generator (ICKG)}
%: knowledge graph construction with Large Language Models}
\label{sec:ickg}

%\subsection{Supervised fine-tuning for KGC}

One of the objectives of this work is to propose an automated and scalable pipeline to extract temporal knowledge graphs from unstructured data sources, such as text. Large language models represent a natural choice for this task. Generative LLMs, while usually proficient in a wide array of tasks related to language, often require customization in more specialized applications, such as knowledge graph construction. This can be achieved via \textit{supervised fine-tuning}, which involves the further training of a pre-trained LLM on a curated dataset that is tailored to the task at hand \cite{nie2024survey}.

For the purposes of this work, we develop the \textit{Integrated Contextual Knowledge Graph Generator (ICKG)}\footnote{The ICKG-v3.2 model is publicly available on the HuggingFace
platform for non-commercial research at \ifthenelse{\boolean{anonym}}{ANONYMISED LINK}{\url{https://huggingface.co/victorlxh/ICKG-v3.2}.}}, an open-sourced fine-tuned LLM, % based on the Mistral-7B model, 
which is optimised for knowledge graph construction tasks and uses the GPT-4 API for data generation. The training workflow of ICKG was divided into the following steps:

\begin{enumerate}
    \item First, a fine-tuning dataset is constructed from a small set of $5,000$ open-sourced financial news articles. These are passed to GPT-4 one-by-one with a knowledge graph extraction prompt giving detailed instructions on the required output type, consisting in triplets extracted from the article. Additionally, our prompt asks to classify entities into a pre-defined set of \textit{categories}, or meta-entities. % related to financial concepts.
    \item Next, an additional data quality filter is applied to the resulting output. Only responses that strictly adhere to the instruction prompt and %yielded a sufficiently rich set of quadruples (more than 5 per article) 
    return more than 5 quadruples per article were retained. This helps reducing the effect of noise and randomness in the GPT-4 output, refining the quality of the quadruples beyond the native capabilities of GPT-4. %In particular, data points generating fewer than five triplets per article were excluded.
    \item The resulting set of quadruples is used to fine-tune the open-sourced Mistral 7B model \citep{Jiang23}, obtaining the final Integrated Contextual Knowledge Graph Generator (ICKG). The fine-tuning process was conducted over approximately 10 hours, utilizing 8 A100 GPUs with 40GB memory each. 
\end{enumerate}

\begin{figure}
  \centering
  \includegraphics[width=\linewidth]{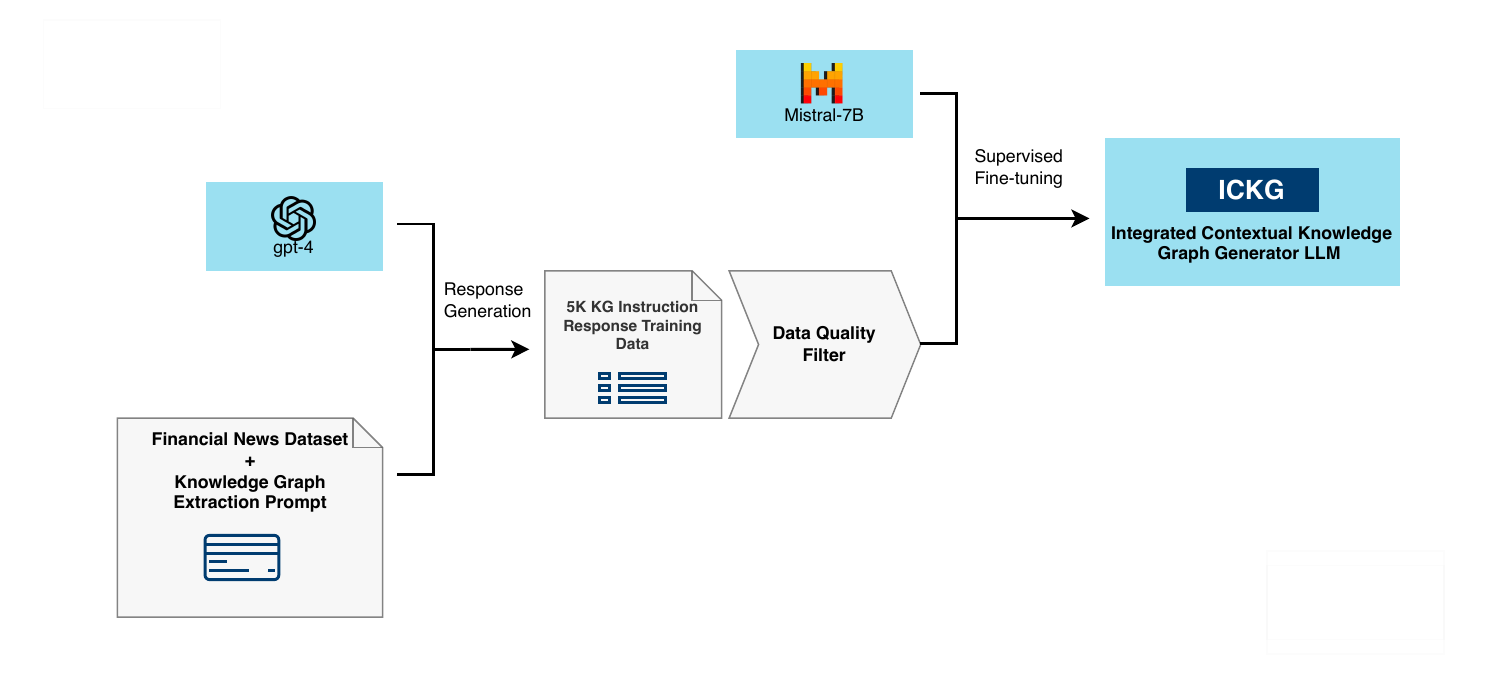}
  \caption{Flowchart of the fine-tuned ICKG LLM for knowledge graph construction, outlining the training methodology.}
  \label{fig:ickg-train}
\end{figure}

%\subsection{Data}

%This study utilizes multiple Temporal Knowledge Graphs (TKGs) datasets, encompassing both widely-recognized benchmark datasets and our novel Financial Dynamic Knowledge Graph (FinDKG) curated from Wall Street Journal full news corpus. We also develop a instruction-following demonstrations dataset to fine-tune our LLM.

%\subsection{Knowledge Graph Construction Instruction Dataset}

%We develop an instruction-following demonstrations dataset on knowledge graph construction task leveraging the GPT-4 API. The dataset is used for fine-tuning the open-source LLM to derive the ICKG model.

%We initiated the data collection with a sample of 5,000 news articles. Subsequent to generating triplet responses via the GPT-4 API, we instituted a second layer of data quality filter. 
%Only responses that strictly adhered to the instruction prompt and yielded a sufficiently rich set of triplets were retained. Specifically, data points generating fewer than five triplets per article were excluded. This data filtering strategy further refined the quality of our supervised fine-tuning dataset, even above the native capabilities of GPT-4, thereby promoting enhanced triplet extraction in the resulting LLM.

The full workflow is depicted in Figure~\ref{fig:ickg-train} diagram.
Figure~\ref{fig:ickg} displays an example of this pipeline, where an open-access news article is passed as input to the LLM, describing a set of predefined entity categories and relations and required output type. The output of the procedure is a set of quintuples representing the resulting KG.

\subsection{The Financial DKG (FinDKG) dataset}
\label{sec:findkg}

Open-source real-world knowledge graphs are relatively scarce, %. This is 
particularly in the financial sector. %, where the cost of curating accurate, reliable, and timely data is prohibitive. To bridge this gap, our study takes on the ambitious task of constructing a global 
Therefore, %one additional 
a contribution of this article is to provide an open-sourced financial dynamic knowledge graph dataset, called FinDKG\footnote{FinDKG is available to download at \ifthenelse{\boolean{anonym}}{ANONYMISED LINK}{\url{https://xiaohui-victor-li.github.io/FinDKG/\#data}.}}, constructed from scratch utilising our ICKG LLM proposed in the previous section.
We collected approximately 400,000 financial news articles from the Wall Street Journal via open-source web archives, spanning from 1999 to 2023. Each article includes metadata such as release time, headlines, categories, in addition to the full textual content. We excluded articles with themes not closely related to economics and finance (such as entertainment, book recommendations, opinion columns). %Additionally, we excluded opinion columns. 

\begin{figure*}[h]
  \centering
  \includegraphics[width=0.9\linewidth]{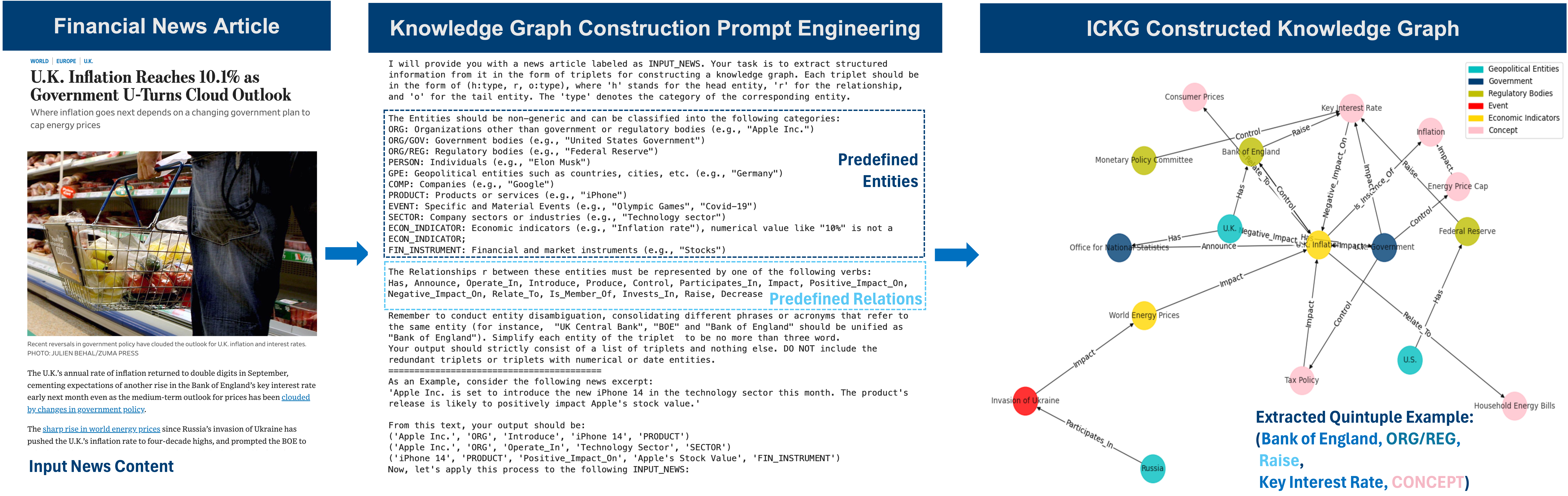}
  \caption{Illustration of the ICKG-enabled knowledge graph generation pipeline for FinDKG, representing the conversion of textual news articles into structured dynamic knowledge graph quintuples.}
  \label{fig:ickg}
\end{figure*}

%This study introduces the Financial Dynamic Knowledge Graph (FinDKG), a daily-resolved, temporal knowledge graph generated from curated global financial news. 
ICKG is used to extract quintuples consisting in entities, entity categories, and relation type from each news article, with timestamps corresponding to the release date. The possible relations are restricted to 15 types relevant to financial news, summarised with examples in Table~\ref{tab:relations_findkg}. The entities are tagged with a \textit{category} selected from the list in Table~\ref{tab:categories_findkg}. Additionally, the resulting quintuples undergo entity disambiguation via Sentence-BERT \citep{reimers2019sentence,zeakis2023pre}. An example of this procedure is given in Figure~\ref{fig:ickg}. 

 \begin{table*}
 \centering
 \scalebox{0.65}{
   \begin{tabular}{lll}
     \toprule
\textbf{Relation} & \textbf{Definition} & \textbf{Example} \\
        \midrule
        Has & Indicates ownership or possession, often of assets or subsidiaries in a financial context. & Google Has Android \\
        Announce & Refers to the formal public declaration of a financial event, product launch, or strategic move. & Apple Announces iPhone 13 \\
        Operate In & Describes the geographical market in which a business entity conducts its operations. & Tesla Operates In China \\
        Introduce & Denotes the first-time introduction of a financial instrument, product, or policy to the market. & Samsung Introduces Foldable Screen \\
        Produce & Specifies the entity responsible for creating a particular product, often in a manufacturing or financial product context. & Pfizer Produces Covid-19 Vaccine \\
        Control & Implies authority or regulatory power over monetary policy, financial instruments, or market conditions. & Federal Reserve Controls Interest Rates \\
        Participates In & Indicates active involvement in an event that has financial or economic implications. & United States Participates In G20 Summit \\
        Impact & Signifies a notable effect, either positive or negative, on market trends, financial conditions, or economic indicators. & Brexit Impacts European Union \\
        Positive Impact On & Highlights a beneficial effect on financial markets, economic indicators, or business performance. & Solar Energy Positive Impact On ESG Ratings \\
        Negative Impact On & Underlines a detrimental effect on financial markets, economic indicators, or business performance. & Covid-19 Negative Impact On Tourism Sector \\
        Relate To & Points out a connection or correlation with a financial concept, sector, or market trend. & AI Relates To FinTech Sector \\
        Is Member Of & Denotes membership in a trade group, economic union, or financial consortium. & Germany Is Member Of EU \\
        Invests In & Specifies an allocation of capital into a financial instrument, sector, or business entity. & Warren Buffett Invests In Apple \\
        Raise & Indicates an increase, often referring to capital, interest rates, or production levels in a financial context. & OPEC Raises Oil Production \\
        Decrease & Indicates a reduction, often referring to capital, interest rates, or production levels in a financial context. & Federal Reserve Decreases Interest Rates \\
  \bottomrule
  \end{tabular}
  }
  \caption{Relation types in the FinDKG dataset.}
  \label{tab:relations_findkg}
 \end{table*}

\begin{table}
\centering
\scalebox{0.625}{
   \begin{tabular}{lll}
     \toprule
\textbf{Category} & \textbf{Definition} & \textbf{Example} \\
        \midrule
        ORG & Non-governmental and non-regulatory organisations. % or regulatory bodies. 
        & \ifthenelse{\boolean{anonym}}{ANONYMISED INSTITUTION}{Imperial College London} \\
        ORG/GOV & Governmental bodies. & UK Government \\
        ORG/REG & Regulatory bodies. & Bank of England \\
        GPE & Geopolitical entities like countries or cities. & United Kingdom \\
        PERSON & Individuals %people often 
        in influential or decision-making roles. & Jerome Powell \\
        COMP & Companies across sectors. & Apple Inc. \\
        PRODUCT & Tangible or intangible products or services. & iPhone \\
        EVENT & Material events with financial or economic implications. & Brexit \\
        SECTOR & Sectors or industries in which companies operate. & Technology Sector \\
        ECON IND & Non-numerical indicators of economic trends or states. & Inflation Rate \\
        FIN INST & Financial and market instruments. & S\&P 500 Index \\
        CONCEPT & Abstract ideas, themes, or financial theories. & Artificial Intelligence \\
  \bottomrule
  \end{tabular}
  }
  \caption{Entity categories in the FinDKG dataset.}
  \label{tab:categories_findkg}
 \end{table}

% Figure~\ref{fig:ickg} displays an example of the proposed pipeline, where an open-access news article is passed as input to the ICKG LLM, describing a set of predefined entity categories and relations and required output type. The LLM is also asked to perform \textit{entity disambiguation}, grouping together entities having the same meaning. The output of the procedure is a set of quadruples which represent the resulting knowledge graph, equipped with the timestamps associated with the input source (for example, a news article or an image). Using this %serves as a promising data processing engine for our framework provides an automated method for extracting TKGs from heterogeneous data sources in scalable fashion.

\begin{figure}[h]
  \centering
\includegraphics[width=.95\linewidth]{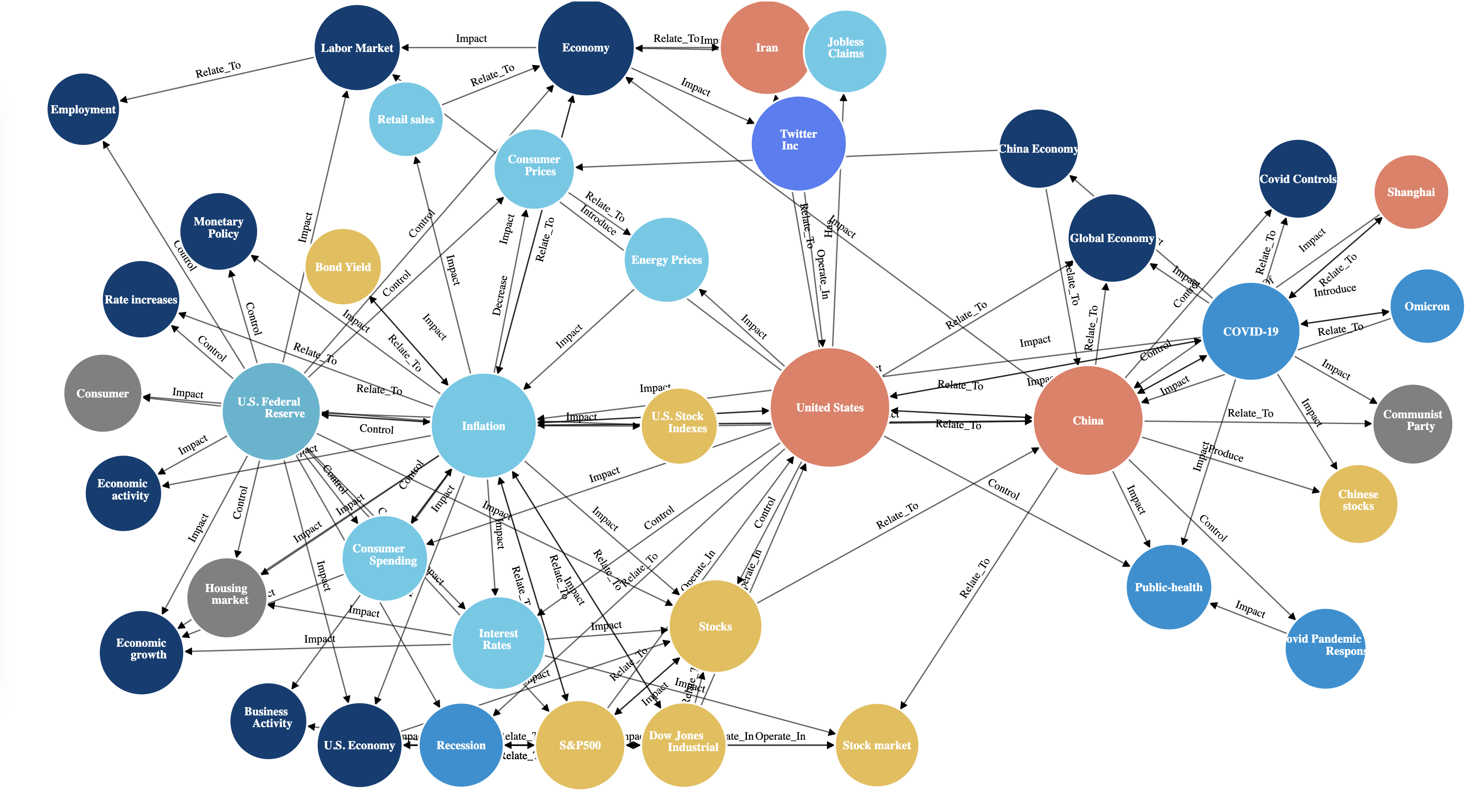}
  \caption{Subgraph of FinDKG’s most influential entities as of January 1, 2023. Entities are coloured by category.}
  \label{fig:fig_KGsnapshot}
\end{figure}

Figure~\ref{fig:fig_KGsnapshot} presents a snapshot subgraph of FinDKG as of January 2023, highlighting the most relevant entities at the time, ranked by graph centrality metrics. %This visualization demonstrates the graph's potential utility for investigating contemporary issues, such as 
The graph shows signs of the geopolitical tensions between the United States and China, the rising global economic pressure of high inflation, and the effect of the COVID-19 pandemic. % and its variants. 
The resulting dataset is used in Section~\ref{sec:experiments} for testing the graph learning procedure for DKGs proposed in Section~\ref{sec:learning}.

%For TKG model training and evaluation, a customized FinDKG study dataset is created as shown in Table~\ref{tab:tab_TKGs}, adopting the conventional temporal knowledge graph dataset structure, including train/validation/test splits organized chronologically. The event triplets are aggregated at weekly intervals to serve as the basic unit of time. 

\section{Graph Learning via KGT\MakeLowercase{ransformers}}
\label{sec:learning}

Dynamic knowledge graph learning %is a fundamental task to capture the inherent 
consists in the task of estimating a model which captures the
structural and temporal characteristics of the observed data. %Recent advancements in dynamic knowledge graph learning methodologies concentrate predominantly on 
%Within the context of DKGs, models proposed in the literature are built for solving two core tasks: % problem setups: DKG 
%\textit{interpolation} and \textit{extrapolation}. Interpolation-based methods are aimed at filling in missing facts for given time intervals, %frames $t : (0 \leq t \leq T)$, 
%whereas extrapolation-based methods predict future facts beyond the known time horizon. % $t : ( t \geq T)$.
The focus of this work is the \textit{extrapolation} task, aimed at predicting future facts beyond the known time horizon, particularly \textit{link prediction}: given a DKG $\mathcal G_t$, source entity $s$, a relation $r$, and a future time $t$, the objective is to predict the most likely object entity $o^\ast$ which will complete the connection, forming the quadruple $(s, r, o^\ast, t)$. More formally, for each triplet $(s,r,t),\ s\in\mathcal E,\ r\in\mathcal R,\ t\in\mathbb R_+$, the objective is to estimate ranking functions %$f_{(s,r,t)}:\mathcal E\to\mathbb{R}$ 
expressing the likelihoods of quadruples $(s,r,o,t),\ o\in\mathcal E$ to occur, as a function of $o\in\mathcal E$. In this work, we %propose to 
learn these functions via %a novel GNN architecture called 
the novel %\textit{Knowledge Graph Transformer (KGTransformer)}, 
\textit{KGTransformer},
described in the next section. 

%This prediction task naturally aligns with a ranking framework. Formally, given the historical graph \( \mathcal{G}_{< t} \), we aim to rank potential future links \( (s, r, ?, t) \) based on their likelihood. The ranking function \( f \) can be mathematically framed as:

%\begin{equation}
%f: (s, r, \mathcal{O}, t, \mathcal{G}_{< t}) \rightarrow \mathbb{R}
%\end{equation}

%where \( \mathcal{O} \) represents the set of all possible object entities, and \( f \) assigns a real-valued likelihood score to each candidate link. The higher the score, the more likely the link. The extrapolation problem setup, despite its challenges, is intriguing due to its potential to forecast future events, thereby holding immense practical value for real-world dynamic KG applications.

\subsection{The Knowledge Graph Transformer} \label{sec:kgtransformer}

In this section, we introduce the KGTransformer, an attention-based graph neural network (GNN) designed to construct lower dimensional representations of the entities, called \textit{graph embeddings}. In addition to standard GNN architectures, KGTransformer incorporates meta-entities via an extended graph attention mechanism based on \cite{hu2020heterogeneousgraphtransformer}, %providing refined embedding representations of the entities, 
borrowing strength across %pre-existing 
entity categories. %Figure~\ref{fig:kgtransformer} shows the model architecture of the proposed KGTransformer.

%\begin{figure}
%  \centering
%  \includegraphics[width=\linewidth]{charts/KGTransformer_diagram.pdf}
 % \caption{Diagram of the KGTransformer model architecture.}
%  \label{fig:kgtransformer}
%\end{figure}

Consider a KG $\mathcal{G} = (\mathcal{E}, \mathcal{R}, \mathcal{F})$, where $N=\vert\mathcal{E}\vert$. The KGTransformer %utilises $L\in\mathbb N$ 
layer %, characterised by 
produces %a latent feature representation 
an embedding $Y^{(\ell)}\in\mathbb R^{N\times D_\ell}$ of the entities, where $D_\ell\in\mathbb N$ is the latent dimension of the $\ell$-th layer, for $\ell\in\{1,\dots,L\}$, initialised from a latent representation $Y^{(0)}\in\mathbb R^{N\times D_0},\ D_0\in\mathbb{N}$. 
 %, with entities \( \mathcal{E} \), relationships \( \mathcal{R} \). Entity \( i \) is initialized with trainable embeddings \( Y^{0} \).
 The latent features $Y^{(\ell)}$ obtained as output of the $\ell$-th layer are passed as input to the $(\ell+1)$-th layer of the full network architecture, until a final output $Y^{(L)}\in\mathbb R^{N\times D}$ is obtained. 
%Let \( Y^{(l)} \in \mathbb{R}^{N \times D } \) denote the output of the \(l\)-th KGTransformer layer, where \( N \) represents the number of nodes and \( D \) signifies the dimension of the node features at layer \( l \). This serves as the input for the succeeding \((l+1)\)-th layer. After sequentially applying \( L \) such layers, we arrive at \( Y^{(L)} \), a contextualized representation of the graph. As an extension of the Graph Attention Network (GAT) \cite{velivckovic2017graph}, the core functionality of KGTransformer is formulated as:

At the $\ell$-th layer, the latent features $Y^{(\ell)}\in\mathbb N^{N\times D_\ell}$ consist in an aggregation operation between $H\in\mathbb N$ sub-vectors of the form $Y^{(\ell)}\in\mathbb N^{N\times D_{\ell,h}},\ D_{\ell,h}\in\mathbb N$, where $\sum_{h=1}^H D_{\ell,h}=D_{\ell}$, such that:
\begin{equation}
    Y^{(\ell)} = \left[Y_1^{(\ell)} , \dots, Y^{(\ell)}_H \right]\in\mathbb R^{N\times D_{\ell}},
\end{equation}
by concatenation.
Each component refers to a part of the input from the previous layer, creating a so-called \textit{multi-head} system \cite{vaswani2017attention}.

At the $\ell$-th layer, the basic update function for latent features $Y^{(\ell)}_h[o]$ for an entity $o\in\mathcal E$ in the KGTransformer consists in combination between the so-called \textit{message vectors}, weighted by \textit{attention scores}, according to the following aggregation equation:% based on multi-head attention and message vectors:
\begin{equation}
Y^{(\ell)}_h [o] = \psi\left(
\sum_{s\in\mathcal{E},r\in\mathcal{R}:s\in\mathcal{N}_r(o)} \text{Atn}^{(\ell)}_{h}(s, r, o)~\text{Msg}^{(\ell)}_{h}(s, r, o)\right),
\label{eq:kgtransformer}
\end{equation}
%{\color{red} Check that all formulas are correct in terms of how they have been rewritten. Add references where needed.}
where %$\oplus$ denotes an aggregation operation (in this case summation),
%$\odot$ is the Hadamard product,
$\mathcal N_r(o)=\{s\in\mathcal E: (s,r,o)\in\mathcal F\}$ is the set of type-$r$ neighbours for the entity $o$, and $\text{Atn}^{(\ell)}_{h}(\cdot)\in\mathbb R$ and $\text{Msg}^{(\ell)}_{h}(\cdot)\in\mathbb R^{D_{\ell,h}}$ are attention and message vectors, % respectively, 
calculated %as a function of the embeddings 
from $Y^{(\ell-1)}$. % from the $(\ell-1)$-th layer. 
%\begin{equation}
%Y^{(l+1)}[o] =  \text{Aggregate} \big( \text{Attention}(s, o) \times \text{Message}(s) \big)
%\end{equation}
Additionally, $\psi(\cdot)$ %$%=\max\{0.01\cdot x , x\}$ 
is the element-wise %leaky rectified linear unit (
Leaky-ReLU activation function. 

%In this equation, \( \textit{Attention} \) computes the mutual attention between source nodes \(s\) and the object node \(o\), \( \textit{Message} \) computes the informational content emanating from the source node \(s\), and \( \textit{Aggregate} \) consolidates these messages in a weighted manner based on the attention scores. This representation is trainable and is subsequently employable for various downstream graph-oriented learning tasks.

%\textbf{Meta-Relation:} The KGTransformer introduces an extended attention mechanism that incorporates meta-relations, defined as triplets \( \langle \tau(s), \phi(r), \tau(o) \rangle \), where \(s\) and \(o\) are the source and object nodes, and \(r\) is the relation between them. The meta-relation captures not only the relationship type but also the types of the nodes involved.

\paragraph{KGTransformer attention vectors.} 
The KGTransformer attention scores $\text{Atn}^{(\ell)}_{h}(s, r, o)$ in \eqref{eq:kgtransformer} are calculated by applying the softmax transformation (denoted $\sigma$) on a concatenation of scores $\alpha^{(\ell)}_{h}(s, r, o)$ across entities in neighbourhoods $\mathcal{N}_r(o)$ for each relation $r\in\mathcal R$:
\begin{equation}
    \text{Atn}_h^{(\ell)}(s, r, o) = 
    \sigma\left(\left[\big\|_{s\in\mathcal{E},r\in\mathcal{R}:s\in\mathcal{N}_r(o)} \alpha^{(\ell)}_{h}(s, r, o) \right] \right),
    %%\sigma\left(\left[\alpha^{(\ell)}_{1}(s, r, o),\dots,\alpha^{(\ell)}_{H}(s, r, o)\right]\right).
    \label{eq:attention_scores}
\end{equation}
where $\|_\cdot$ denotes the concatenation operator.
The normalisation via the softmax ensures that the weights in the update \eqref{eq:kgtransformer} sum to $1$.
%The value of $\text{Atn}_h^{(\ell)}(s,r,o)$ in \eqref{eq:kgtransformer} corresponds to the $h$-th entry of $\text{Atn}^{(\ell)}(s, r, o) $.

Each of the attention scores $\alpha^{(\ell)}_{h}(s,r,o),\ h=1,\dots,H$ in \eqref{eq:attention_scores} is obtained after incorporating \textit{meta-entities}. In particular, we assume that   a function $\tau:\mathcal E\to \mathcal C_{\mathcal E}$ %and $\phi:\mathcal R\to\mathcal C_{\mathcal R}$ 
exists, mapping each entity %and relation 
to an entity type, % and relation type respectively, 
where all possible types are described by %two sets 
the set $C_{\mathcal E}$. % and $C_{\mathcal R}$. 
%The KGTransformer introduces an extended attention mechanism that incorporates meta-relations, defined as triplets \( \langle \tau(s), \phi(r), \tau(o) \rangle \), where \(s\) and \(o\) are the source and object nodes, and \(r\) is the relation between them. The meta-relation captures not only the relationship type but also the types of the nodes involved.
For example, consider the relation \textit{Invent} between the source entity \textit{OpenAI}, which is of type \textit{Company}, and the object entity \textit{ChatGPT}, which is of type \textit{Product}. %The relation between them is \texttt{Invent}. 
In the context of meta-entities, this could be represented as $\tau(\textit{OpenAI})=\textit{Company}$ and $\tau(\textit{ChatGPT})=\textit{Product}$. Meta-entities are incorporated in the %attention mechanism 
architecture via tensors $\mu^{(\ell)}_h\in\mathbb R^{\vert\mathcal C_{\mathcal E}\vert\times\vert\mathcal R\vert\times\vert\mathcal C_{\mathcal E}\vert},\ h=1,\dots,H,\ \ell=1,\dots,L$, following the same approach of \cite{hu2020heterogeneousgraphtransformer} on heterogeneous graphs. Following \cite{hu2020heterogeneousgraphtransformer}, the proposed KGTransformer attention score for the $h$-th head is: 
\begin{equation}
\alpha^{(\ell)}_{h}(s,r,o) = \frac{{K^{(\ell)}_h[s]}^\intercal W^{(\ell)}_{h,r} Q^{(\ell)}_h[o] \cdot \mu^{(\ell)}_h[\tau(s), r, \tau(o)]}{\sqrt{D_{\ell,h}}},
\label{eq:attention}
\end{equation}
%\textbf{KGTransformer Attention:} The attention mechanism is refined by incorporating meta-relations \( \langle \tau(s), \phi(r), \tau(o) \rangle \), where \( \tau \) represents node types and \( \phi \) indicates the relationship type. This allows the model to adaptively scale attention scores using a tensor \(\mu\), enhancing the representational capacity:
%\begin{equation}
%\text{Attn}(s,r,o) = \frac{K(s) W_{\phi(r)} Q(o) \cdot \mu_{\langle \tau(s), \phi(r), \tau(o) \rangle}}{\sqrt{d_k}}
%\end{equation}
%where $D_{\ell,h}$ is the corresponding latent dimension for the $h$-th head, which is often set to $D_\ell/H$, %in the literature, 
%such that $\sum_{h=1}^H D_{\ell,h}=D_\ell$. 
where the vectors $K^{(\ell)}_h[s],\ Q^{(\ell)}_h[o]\in\mathbb R^{D_{\ell,h}\times 1}$ in \eqref{eq:attention} are called \textit{key} and \textit{query} vectors for entities $s$ and $o$, and $W^{(\ell)}_{h,r}\in\mathbb R^{D_{\ell,h}\times D_{\ell,h}}$ is a trainable weighting matrix. %, evaluated at the entry corresponding to $\langle\tau(s), r, \tau(o)\rangle$. % in \eqref{eq:attention}. 
% key and query vectors are common setting in Transformers (attention) 
The key and query vectors are derived from the latent features at the previous layer: %, combined with linear projections specific to the node category: 
\begin{align}
    K^{(\ell)}_h[s] = P^{(\ell)}_{h,\tau(s)}Y^{(\ell-1)}[s], & & 
    Q^{(\ell)}_h[o] = R^{(\ell)}_{h,\tau(o)}Y^{(\ell-1)}[o],
    \label{eq:key_projection}
\end{align}
where $P^{(\ell)}_{h,c},R^{(\ell)}_{h,c}\in\mathbb{R}^{D_{\ell,h}\times D_{\ell-1}},\ c\in\mathcal C_{\mathcal E}$, are trainable matrices.

\paragraph{KGTransformer message vectors.} 
Similarly to the attention scores in \eqref{eq:attention_scores}, message vectors %$\text{Msg}^{(\ell)}_{\text{KGT}}(\cdot)$ in \eqref{eq:kgtransformer} are calculated via concatenation of $H\in\mathbb N$ message heads:
%\begin{equation}
%    \text{Msg}^{(\ell)}_{\text{KGT}}(s,r,o) = \left[\text{Msg}^{(\ell)}_{1}(s,r,o),\dots,\text{Msg}^{(\ell)}_{H}(s,r,o) \right].
%\end{equation}
%Each message head is 
are obtained via different linear projections applied to the embedding $Y^{(\ell-1)}$ from the previous layer \cite{hu2020heterogeneousgraphtransformer}:
\begin{equation}
    \text{Msg}^{(\ell)}_{h}(s,r,o) = %\psi\left(
    Z^{(\ell)}_{h, r}
    M^{(\ell)}_{h, \tau(s)} Y^{(\ell-1)}[s],
    %\right), %W_{\text{MSG}}^{\tau(s), \phi(r)},
\end{equation}
where  %$\psi(\cdot)$ %$%=\max\{0.01\cdot x , x\}$ 
%is the element-wise leaky rectified linear unit (Leaky-ReLU) activation, and 
$M^{(\ell)}_{h,\tau(s)} \in \mathbb R^{D_{\ell,h}\times D_{\ell-1}}, Z^{(\ell)}_{h, r} \in \mathbb R^{D_{\ell,h}\times D_{\ell,h}}$ are %linear projection 
matrices specific to the $h$-th head, meta-entity $\tau(s)$, % for the source entity $s$, 
and relation $r$.

\subsection{Time-evolving updates for DKGs}
\label{sec:time_dkg}

So far, Section~\ref{sec:kgtransformer} only considered the case of a static knowledge graph. In this section, we discuss how to incorporate two different types of time-varying representations, called \textit{temporal} and \textit{structural} embeddings, following the EvoKG framework in \cite{park2022evokg}.

Let $\mathcal{G}_{t}=(\mathcal E,\mathcal R,\mathcal F_t)$ be a DKG observed at discrete time points $t=1,\dots,T$, such as $\mathcal F_t\subseteq\mathcal F_{t^\prime}$ for $t<t^\prime$. We write $\tilde{\mathcal F}_t=\mathcal F_t\setminus \mathcal F_{t-1}$ to denote the set of facts occurring in the time interval between $[t-1,t)$. This representation can be used to construct a set of KGs $\tilde{\mathcal{G}}_{t}=(\mathcal E,\mathcal R,\tilde{\mathcal F}_t)$ where $\tilde{\mathcal F}_t\cap\tilde{\mathcal F}_{t^\prime}=\varnothing$ for $t\neq t^\prime$. 

First, we apply KGTransformer independently on each graph $\tilde{\mathcal G}_t$, obtaining an embedding representation $Y_t^{(\ell)}\in\mathbb{R}^{N\times D_{\ell}}$ via \eqref{eq:kgtransformer}, starting from an input embedding $Y_t^{(\ell-1)}\in\mathbb{R}^{N\times D_{\ell-1}}$:
\begin{equation}
    Y_t^{(\ell)} = \mathrm{KGTransformer}\left(Y_t^{(\ell-1)},\ \tilde{\mathcal G}_t\right).
\end{equation}
The evolution of the embeddings $Y_t^{(\ell)},\ t=1,\dots, T$ over time is modelled via a recurrent neural network (RNN), resulting in: % a sequence of embeddings: 
\begin{equation}
    V_t^{(\ell)} = \text{RNN}\left(Y_t^{(\ell)},\ V_{t-1}^{(\ell)}\right).
\end{equation}
The values $V_t^{(\ell)}\in\mathbb R^{N\times D_\ell},\ t=1,\dots,T$, are called \textit{temporal embeddings}. Following \citep{park2022evokg}, the temporal embeddings for the unique entities appearing in $\mathcal F_{r,t}=\{(s,r^\prime,o,t)\in\tilde{\mathcal F}_t:r^\prime=r\}$ are averaged to obtain a latent representation for the relations $\tilde{Y}^{(\ell)}_t\in\mathbb{R}^{\vert\mathcal R\vert\times D_\ell}$, which is analogously modelled via an RNN, giving a sequence of \textit{temporal relation embeddings} $\tilde V_t^{(\ell)}\in\mathbb R^{\vert\mathcal R\vert\times D_\ell},\ t=1,\dots,T$, where:
\begin{equation}
    \tilde{V}_t^{(\ell)} = \text{RNN}\left(\tilde{Y}_t^{(\ell)},\ \tilde{V}_{t-1}^{(\ell)}\right).
\end{equation}
We denote the rows of ${V}_t^{(\ell)}$ and $\tilde{V}_t^{(\ell)}$ as $v_{i,t}^{(\ell)}$ and $\tilde v_{r,t}^{(\ell)}$, for entity $i$ and relation $r$ respectively.
%Temporal embeddings capture the sequence of events and interactions among entities, denoted by \( t_i \) for entities and \( t_r \) for relations. Specifically, these embeddings at layer \( l+1 \) and time \( t \) are represented as: 
These embedding representations will be used %in the next section 
to model the conditional probability of the arrival time of the triplets $(s,r,o)\in\mathcal E\times\mathcal R\times\mathcal E$, as in the EvoKG framework \citep{park2022evokg}. 
%\begin{align}
%t^{(l+1, t)}_i &= \text{KGTransformer}\left(\mathcal{G}_t\right), \\
%t^{(l+1, t)}_r &= \text{KGTransformer}(t_r, \mathcal{G}_t)
%\end{align}

%They are updated over time through specialized Recurrent Neural Networks (RNNs):

%\begin{equation}
%t^{(l, t)} = \text{RNN}_{\text{temporal}}(t^{(l, t)}, t^{(l, t-1)})
%\end{equation}

In contrast, the conditional probabilities of the triplets given the graph $\mathcal G_t$ will be modelled via the so-called \textit{structural embeddings} \citep{park2022evokg}. %,  \( u_i \) and \( u_r \) reflect evolving structural relationships via separate RNNs:
These are obtained via a similar mechanism as above: the output of the KGTransformer is used within an RNN. Denoting the initial input embedding as $X^{(\ell-1)}_t\in\mathbb{R}^{N\times D_{\ell-1}}$, we write:
\begin{align}
    X_t^{(\ell)} = \mathrm{KGTransformer}\left(X_t^{(\ell-1)},\ \tilde{\mathcal G}_t\right),\ 
    U_t^{(\ell)} = \text{RNN}\left(X_t^{(\ell)},\ U_{t-1}^{(\ell)}\right). \nonumber
\end{align}
The values $U_t^{(\ell)}\in\mathbb R^{N\times D_\ell},\ t=1,\dots,T$, are called \textit{structural embeddings}. As before, averaging over the entities appearing in the sub-graph of type $r\in\mathcal R$ at time $t$ gives embeddings $\tilde{U}_t^{(\ell)}\in\mathbb R^{\vert\mathcal R\vert\times D_{\ell}}$. As before, these are modelled via a recurrent neural network:
\begin{equation}
    \tilde{U}_t^{(\ell)} = \text{RNN}\left(\tilde{X}_t^{(\ell)}, \\ \tilde{U}_{t-1}^{(\ell)}\right). 
\end{equation}
As before, $u_{i,t}^{(\ell)}$ and $\tilde u_{r,t}^{(\ell)}$ are used to denote the rows of $U_t^{(\ell)}$ and $\tilde U_t^{(\ell)}$ respectively, corresponding to the structural embeddings at time $t$ for entity $i$ and relation $r$. 

\subsection{Dynamic knowledge graph learning}

In this section, a probabilistic framework for learning DKGs is discussed, based on the work of \cite{jin2019recurrent,park2022evokg}, integrated with the KGTransformer time-varying embeddings discussed in Section~\ref{sec:time_dkg}. % introduces a probabilistic approach to model the distribution of DKGs, building upon the entity and relation embeddings obtained from KGTransformer. 
The objective of the graph learning procedure is to %model the generative process of the DKG up to time $T$, 
estimate the model parameters that best describe the observed graph $\mathcal G_T$ under the proposed model. %denoted as $p(\mathcal{G}_T)$. 
Using $\tilde{\mathcal G}_1,\dots,\tilde{\mathcal G}_T$, we can decompose the probabilities associated with events occurred in the graph $\mathcal G_T$ as follows: %, it is possible to write:
%\begin{equation}
%    p(\mathcal G_T) = p(\tilde{\mathcal G}_1,\dots,\tilde{\mathcal G}_T) = \prod_{t=1}^T p(\tilde{\mathcal G}_t\mid\mathcal G_{t-1}).
%\end{equation}
%Following \cite{park2022evokg}, we assume that $p(\tilde{\mathcal G}_t\mid\mathcal G_{t-1})$ can be written as a product over the facts occurred in $\tilde{\mathcal F}_t$:
%\begin{equation}
%    p(\tilde{\mathcal G}_t\mid\mathcal G_{t-1}) = 
%    \prod_{(s,r,o,t)\in\tilde{\mathcal F}_t} p(t\mid s,r,o,\mathcal G_{t-1}) \times p(s,r,o\mid\mathcal G_{t-1}).
%    \label{eq:dual_decomp}
%\end{equation}
\begin{align}
p(\mathcal G_T) &= p(\tilde{\mathcal G}_1,\dots,\tilde{\mathcal G}_T) = \prod_{t=1}^T p(\tilde{\mathcal G}_t\mid\mathcal G_{t-1}) \\ 
&=\prod_{t=1}^T\prod_{(s,r,o,t)\in\tilde{\mathcal F}_t} p(t\mid s,r,o,\mathcal G_{t-1})\ p(s,r,o\mid\mathcal G_{t-1}).
    \label{eq:dual_decomp}
\end{align}
The decomposition in \eqref{eq:dual_decomp} partitions the conditional probability into two components: $p(s,r,o \mid \mathcal{G}_{t-1})$ captures the evolving graph structure, whereas $p(t\mid s,r,o, \mathcal{G}_{t-1}) \) controls the temporal dynamics. Therefore, a model should be postulated on both these probabilities to capture both temporal and structural characteristics of DKGs.

\paragraph{Modelling the graph structure.} To approximate $ p(s,r,o \mid \mathcal G_t)$, we use embeddings that represent the time-varying structural components of both entities and relationships. Let $u_{i,t},\tilde u_{r,t}\in\mathbb R^D, D\in\mathbb N$, be the structural embeddings for entity $i$ and relation $r$, updated until time $t$, obtained from the final layer of the KGTransformer. 
Additionally, we combine those into a global embedding $g_t=(g_{t,1},\dots,g_{t,D})\in\mathbb R^D$ that aggregates the %structural 
embeddings %$u_{i,t}$ 
of all entities %in the DKG 
up to time $t$ \citep{jin2019recurrent}. Each entry of $g_t$ is computed as follows:
\begin{equation}
 g_{t,j} = \max_{i \in \mathcal E_t} \left\{ u_{i,t,j} \right\},\ j=1,\dots,D,
\end{equation}
where $\mathcal E_t=\{s\in\mathcal E:(s,r,o)\in\tilde{\mathcal F}_t \vee (o,r,s)\in\tilde{\mathcal F}_t, r\in\mathcal R, o\in\mathcal E\}$ is the set of entities involved in events in $\tilde{\mathcal F}_{t}$. 
The vector $g_t$ is used as a global conditioning variable for computing %event-related 
%conditional probabilities 
$p(s,r,o \mid \mathcal G_{t})$ \cite{jin2019recurrent}.

Following \cite{park2022evokg}, we decompose $p(s,r,o\mid\mathcal G_{t})$ into entity and relationship level components as follows: %. To account for the edge directional nature of the DKG, we propose two separate conditional probability models for both source and object entity:
\begin{equation}
    p(s, r, o \mid \mathcal G_{t}) = p(o \mid \mathcal G_{t}) \times p(r \mid o, \mathcal G_{t}) \times p(s \mid r, o, \mathcal G_{t}). 
    \label{eq:dec}
\end{equation}
Each term is parametrised via a multilayer perceptron (MLP) \citep{park2022evokg}:
\begin{align}
    p(s \mid r, o, \mathcal G_{t}) &= \sigma\left\{\text{MLP}([\tilde u_{r,t}, u_{o,t}, g_t]) \right\}, \\
    p(r \mid o, \mathcal G_{t}) &= \sigma\left\{\text{MLP}([ u_{o,t}, g_t]) \right\}, \\
    p(o \mid \mathcal G_{t}) &= \sigma\left\{\text{MLP}(g_t) \right\}.
    \label{eq:mlps}
\end{align}

Similarly to \eqref{eq:dec}, the equivalent decomposition
\begin{equation}
p(s, r, o \mid \mathcal G_{t}) = p(s \mid \mathcal G_{t}) \times p(r \mid s, \mathcal G_{t}) \times p(o \mid r, s, \mathcal G_{t})
\end{equation}
could also be used, and parametrised via three MLPs as in \eqref{eq:mlps}.

\paragraph{Modelling the temporal dynamics.}  Following \citep{park2022evokg}, we model $p( t \mid s,r,o, \mathcal{G}_{t} ) \) via a mixture of $M\in\mathbb{N}$ log-normal distributions:
\begin{equation}
p( t \mid s,r,o, \mathcal{G}_{t}) = \sum_{m=1}^{M} w_m \phi_{\text{LN}}( t; \mu_m, \sigma_m ),
\end{equation}
where $\phi_{\text{LN}}( t; \mu_m, \sigma_m)$ %represents the 
is the log-normal density function, where $w_m, \mu_m, \sigma_m$ %being 
are the weight, mean, and standard deviation of the $m$-th component, % respectively, 
such that $w_m,\sigma_m\geq 0$ for all $m=1,\dots,M$, and $\sum_{m=1}^M w_m=1$. %The weights \( w_m \) sum to 1. 
Model parameters are learned through an MLP that receives %processes %a %context vector 
inputs composed of concatenated temporal embeddings for each entity and relation derived from the KGTransformer. %, tailored to the specific event \( e \).

\paragraph{Inference on the model parameters.}

The model parameters are learned by minimising a composite loss function, which follows again the approach of \citep{park2022evokg} with a minor adjustment for relational symmetries. In particular, we let the loss function be:
\begin{align}
    \mathcal L =& -\sum_{t=1}^T \sum_{(s,r,o,t)\in\tilde{\mathcal G_t}}\bigg\{\ \lambda_1\log p(t\mid s,r,o, \mathcal G_{t-1})\ \\ &+ \lambda_2\big[ \log p(o\mid \mathcal G_{t-1}) + \log p(r\mid o,\mathcal G_{t-1}) + \log(s\mid r,o,\mathcal G_{t-1})\ \notag \\ 
    &+\log p(s\mid \mathcal G_{t-1}) + \log p(r\mid s,\mathcal G_{t-1}) + \log(o\mid r,s,\mathcal G_{t-1}) \big] \bigg\},
\end{align}
where $\lambda_1,\lambda_2\in\mathbb R_+$ are tunable hyperparameters. In order to manage computational and memory requirements, truncated backpropagation through time \citep[TBPTT; see][]{williams1990efficient, park2022evokg} is used to minimise $\mathcal L$. 

\paragraph{Link prediction.} As described in the introduction, the model performance is evaluated on link prediction, aimed at predicting the most likely object $o$ for an incomplete quadruple $(s, r, ?, t)$. 
The predicted %object 
entity $\hat o$ is obtained as %follows: 
%\begin{equation}
$
  \hat o = \argmax_{o\in\mathcal E} p(o\mid s,r,\mathcal G_t),
  %p_{\text{obj}}(q') = -\log p(o' | s, r, \mathcal{G}_{<t}) = \text{MLP}([u_s^* \parallel u_r^* \parallel g^*]),
$
%\end{equation}
where the distribution $p(o\mid s,r,\mathcal G_t)$ is estimated %under the model 
via the MLP in \eqref{eq:mlps}.  

\section{Experiments and Applications}% and applications}
\label{sec:experiments}

In this section, we test the performance of KGTransformer for link prediction tasks on popular benchmarks used in the literature and on the newly created FinDKG dataset. Additionally, we evaluate the performance of FinDKG, generated by ICKG LLM, in detecting financial trends from the news articles by analysing graph centrality measures. We also explore its application for thematic investing.

\subsection{Link prediction on real-world DKGs} \label{sec:lp}

We conduct experiments on various real-world knowledge graph datasets to evaluate the efficacy of our proposed KGTransformer model, focusing on its performance for link prediction. %, particularly focusing on the extrapolative temporal link prediction task on DKGs.

\paragraph{Performance metrics.} %The primary metric for our evaluation is temporal link prediction, which predicts the likelihood of relationships between entities at future timestamps. 
Following existing literature \citep[see, for example,][]{park2022evokg}, we measure the model's accuracy for link prediction using Mean Reciprocal Rank (MRR) and Hits@n (specifically Hits@3 and Hits@10). %These metrics reflect the model's ability to rank true future links highly among predictions. 
The MRR is defined for a set $\mathcal Q$ of test quadruples by summing the inverses of the ranks associated with each quadruple:
%\[
%\text{MRR} = \frac{1}{|\mathcal{Q}|} \sum_{q \in \mathcal{Q}} \frac{1}{\text{rank}_q},
%\]
$\text{MRR} = \sum_{q \in \mathcal{Q}} \text{rank}_q^{-1} / {|\mathcal{Q}|},$
where \(\text{rank}_q\) is the position of the true link in the ranked list of predictions. 
On the other hand, \textit{Hits@n} measures the proportion of true links ranked within the top-$n$ predictions. %Predictions are tested on timestamps beyond the training phase. % to ensure generalization to unseen data. 
%Additionally, a 
A validation set is used to implement an early stopping mechanism to avoid overfitting. %mitigate potential overfitting.

\paragraph{Baseline models for comparisons.} We compare the performance of the proposed KGTransformer against the following methods: %traditional static and temporal knowledge graph:
\begin{itemize}
  \item Static graph models: %Models like 
  R-GCN \cite{schlichtkrull2018modeling}, which treats the graph as time-invariant, providing a baseline. %by ignoring temporal dynamics.
  \item Temporal graph models: RE-Net \cite{jin2019recurrent} and EvoKG \cite{park2022evokg}. %,  which are designed for temporal edge prediction. %, showcase the advancements in temporal graph analytics.
  \item %KGTransformer variants -- %To validate the importance of incorporating meta-relation types, we also 
  %We contrast KGTransformer with a version
  A KGTransformer version excluding meta-relations (denoted \textit{``KGTransformer w/o node types''} in plots).
\end{itemize}

\paragraph{Implementation details.} The KGTransformer is implemented with two layers of transformation blocks, with each embedding having a dimensionality of 200. We adhere to the original specifications for baseline KG models. All models are optimized using the AdamW algorithm \citep{AdamW} with a learning rate of \(5 \times 10^{-4}\) and an early stopping mechanism triggered after 10 epochs of no validation improvement.

Both model training and evaluations are consistently conducted on an identical computational environment: %, specifically 
a single NVIDIA A100 GPU cloud server %, equipped 
with 40GB of memory. To account for the inherent variability in %deep learning graph 
model training, we employ three distinct random seeds, shared across different models. %\footnote{The selected random seeds are 0, 1, and 41.}. 
The final results are reported as averages over these %different 
training runs. Results across different seeds exhibit minimal variance for the datasets used in this work.

\paragraph{Datasets for evaluation.} We evaluate the performance of the proposed KGTransformer architecture on publicly accessible real-world DKGs used as benchmarks in the literature \citep{park2022evokg}, alongside the FinDKG introduced as part of this work, described in Section~\ref{sec:findkg}. Summary statistics about these datasets are described in Table~\ref{tab:tab_DKGs}. 
ICEWS consists in a collection of cooperative or hostile actions between socio-political actors corresponding to individuals, groups, sectors and nation states \citep{ICEWS_Reference}.  
YAGO dataset is a large, automatically generated knowledge base that combines relational data from Wikipedia and WordNet \citep{YAGO_Reference}, whereas WIKI is based on triplets extracted from the Wikidata database \citep{Wiki_Reference, leblay2018deriving}. 

It must be remarked that the only dataset containing meta-entities is FinDKG: therefore, we expect the benefits of KGTransformer to be particularly evident for this dataset. For the other benchmarks, % datasets, 
the identity mapping function $\tau(s)=s$ is used, implying that $\mathcal E=\mathcal C_{\mathcal E}$. 

\begin{table}[t]
    \centering
    \scalebox{0.8}{
    \begin{tabular}{ lcccccc }
        \toprule
        Dataset & $N_\text{train}$ & $N_\text{val}$& $ N_\text{test}$ & $\vert\mathcal E\vert$ & $\vert\mathcal R\vert$ & $\vert\mathcal C_{\mathcal E}\vert$ \\
        \midrule
        YAGO & 161,540 & 19,523 & 20,026 & 10,623 & 10 & - \\
        WIKI & 539,286 & 67,538 & 63,110 & 12,554 & 24 & - \\
        ICEWS14 & 275,367 & 48,528 & 341,409 & 12,498 & 260 & - \\
        \hline
        FinDKG & 119,549 & 11,444 & 13,069 & 13,645 & 15 & 12 \\
        \bottomrule
    \end{tabular}
    }
    \caption{%Descriptive statistics 
    Summaries of the %real-world 
    DKGs %and FinDKG dataset employed 
    used for model evaluation.} %Interval refers to the minimum temporal spacing between two adjacent events.}
    \label{tab:tab_DKGs}
\end{table}

\begin{figure}[t]
  \centering
  \includegraphics[width=.95\linewidth]{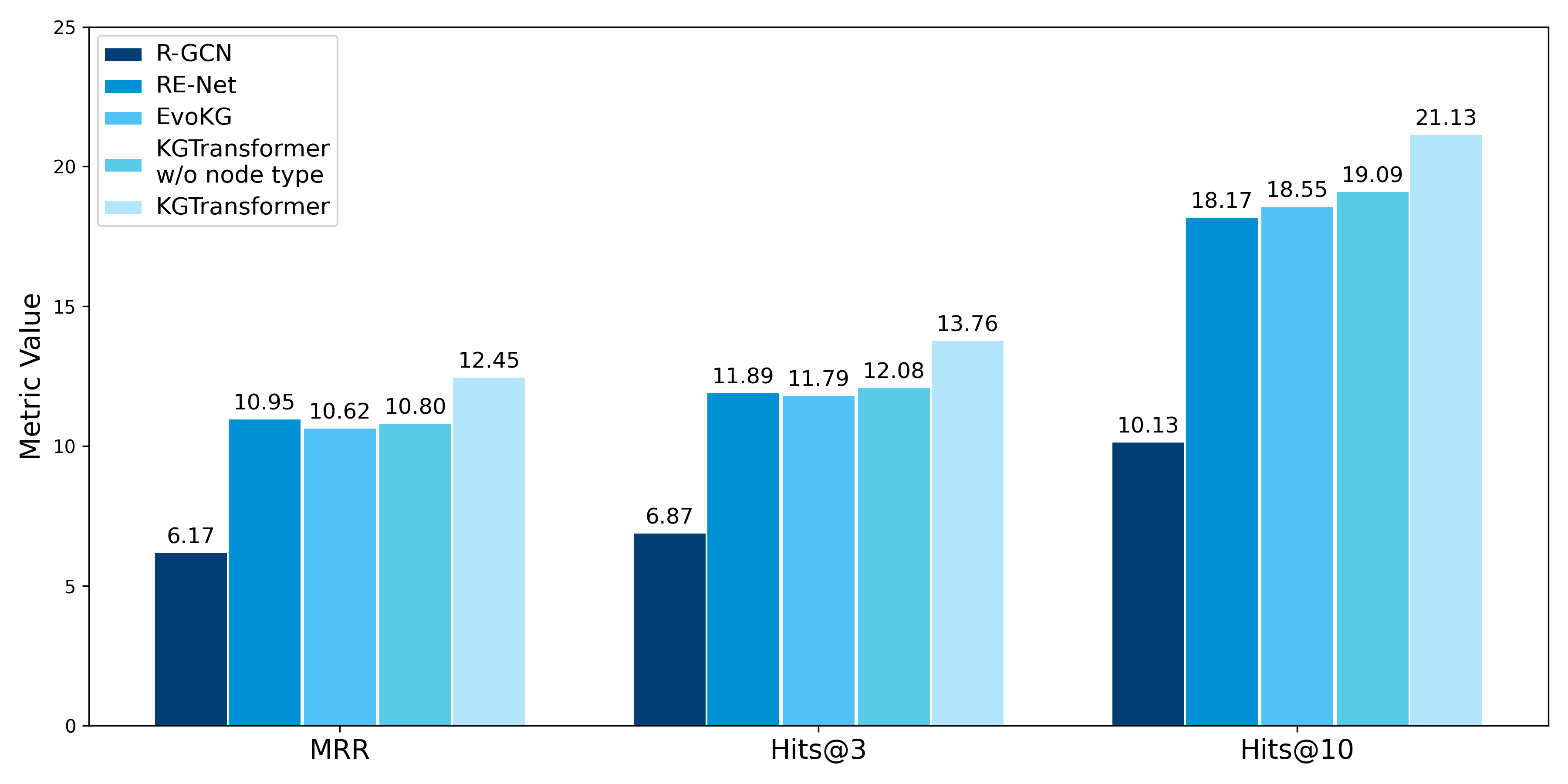}
  \caption{Performance comparison of models on %the 
  FinDKG.} % dataset, evaluated on MRR and Hits@3,10}
  \label{fig:findkg_main}
\end{figure}

\paragraph{Results on benchmarks and FinDKG}
Table~\ref{tab:tab_benchmarkmodel} displays the temporal link prediction scores across the %five 
benchmark DKGs, and Figure~\ref{fig:findkg_main} depicts the results on FinDKG. From the table, it can be seen that the static method R-GCN under-performs in temporal settings, highlighting the importance of temporal features. KGTransformer outperforms competitors on the YAGO and WIKI datasets, but it does not improve performance on %GDELT and 
the ICEWS14 dataset. The advantages of the KGTransformer are more evident on the FinDKG, which explicitly contains entity types (\textit{cf.} Table~\ref{tab:categories_findkg}, \ref{tab:tab_DKGs}). Integrating these types into the KGTransformer enhances performance significantly, resulting in an approximate 10\% improvement in MRR and Hits@3,10 metrics over temporal baselines. This demonstrates the superior performance of KGTransformer when entity categories are also available, providing a way to directly incorporate them into the model architecture. It must be remarked that, when entity categories are not included within the architecture (\textit{“KGTransformer w/o node types”}), the results align closely with the temporal baselines, demonstrating the benefit of introducing this information. % for the purposes of link prediction.

\begin{table}[t]
    \centering
    \scalebox{0.6}{%
    \large
    \begin{tabular}{l|ccc|ccc|ccc}
        \toprule % simulate a bold line by using two hlines
        & \multicolumn{3}{c|}{\textbf{YAGO}} & \multicolumn{3}{c|}{\textbf{WIKI}} & \multicolumn{3}{c}{\textbf{ICEWS14}} \\
        %\cline{2-16}
        \textbf{Model} & \textbf{MRR} & \textbf{H@3} & \textbf{H@10} & \textbf{MRR} & \textbf{H@3} & \textbf{H@10} & \textbf{MRR} & \textbf{H@3} & \textbf{H@10} \\
        \midrule
        R-GCN & 27.43 & 31.24 & 44.75 & 13.96 & 15.75 & 22.05 & 15.03 & 16.12 & 31.47 \\
        RE-Net & 46.35 & 51.93 & 61.47 & 31.45 & 34.23 & 41.15 & 23.81 & 26.57 & \bfseries 42.62 \\
        EvoKG & 49.86 & 57.69 & 65.42 & 42.56 & 47.18 & 52.34 & \bfseries 24.24 & \bfseries 27.25 &  41.97 \\
         %\hline % Increase row height for the following row
         \bfseries KGTransformer & \bfseries 51.33 & \bfseries 59.22 &  \bfseries 67.15 & \bfseries 44.32 & \bfseries 49.27 & \bfseries 53.81 & 23.98 & 26.89 & 41.22 \\
        \bottomrule % simulate a bold line by using two hlines
    \end{tabular}%
    }
    \caption{Performance comparison on the %five 
    benchmark DKGs datasets in terms of MRR, Hits@3,10. Best results are in bold.}
    \label{tab:tab_benchmarkmodel}
\end{table}

\subsection{Trend identification in financial news}

Analysing the results %of KGTransformer 
of FinDKG gives a way to dynamically track the global financial network and evaluate the performance of the ICKG LLM to extract valuable information from financial news. %, %thus enabling the 
%tracking of significant trends influencing the financial system via news reporting. 
To visualise this, we form a series of FinDKGs where rolling 1-month snapshot knowledge graphs were assembled every week on Sundays. These graphs stored the event quadruples of the preceding month. %, offering a fresh perspective on contemporary financial system.
Four graph metrics of centrality were used to quantify the significance of an entity within each temporal knowledge graph: 
degree centrality, betweenness centrality, eigenvector centrality and PageRank. 
%\begin{itemize}
%    \item \textbf{Degree Centrality:} Reflects the number of direct connections an entity has.
%    \item \textbf{Betweenness Centrality:} Measures an entity’s role as an intermediary within the network, highlighting potential influence over systemic risks.
%    \item \textbf{Eigenvector Centrality:} Scores entities based on the influence of their connections, identifying key players connected to other significant entities.
%    \item \textbf{PageRank:} Assesses the importance of an entity based on its links, akin to Eigenvector Centrality but emphasizing link quality over quantity.
%\end{itemize}
To standardize these measures over time for comparability, we apply a rolling one-year $z$-score normalization, making centrality metrics comparable across different times and entities. %Finally, we combine the four standardized centrality indices into a composite KG-based centrality index. 

\begin{figure}[t]
  \centering
  \includegraphics[width=.95\linewidth]{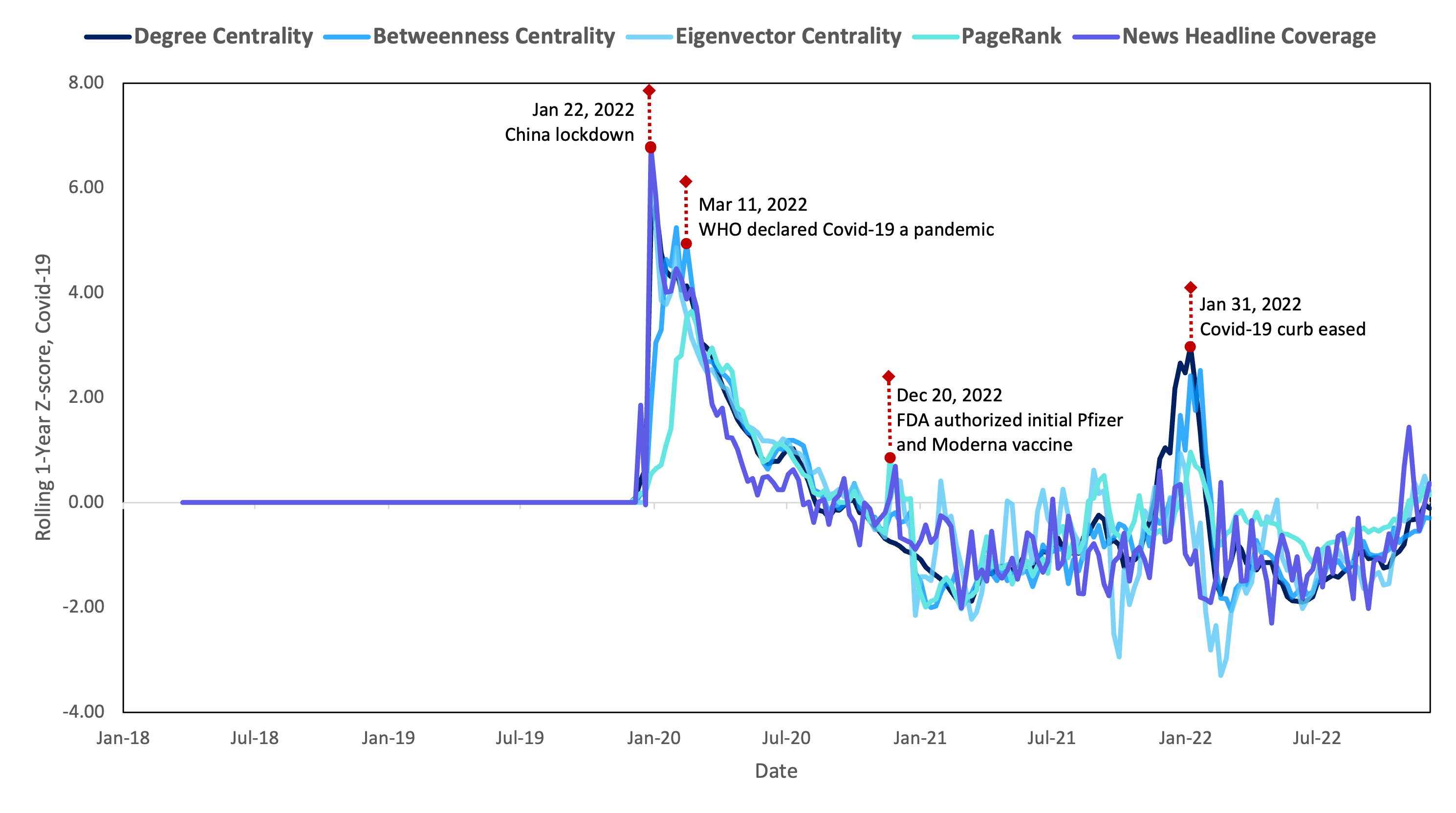}
  \caption{Evolution of the \texttt{Covid-19} entity centrality measures over time between January 2018 and December 2022.} % over time. All indices are sampled at weekly frequency and transformed using rolling 1-year z-score standardization.}
  \label{fig:covid_trend}
\end{figure}

We select the global COVID-19 pandemic as a case study. Figure~\ref{fig:covid_trend} depicts the centrality metrics related to the \texttt{Covid-19} entity as inferred by FinDKG. 
We compare the results with a standard measure based on %news
headline coverage of the topic, commonly used in financial NLP applications \cite{baker2016measuring}. %, serves as a comparison metric. 
%This news coverage metric considers a set of relevant entities as topics and evaluates their significance based on the frequency of news headlines mentioning them within a given timeframe. For the topic \textit{Covid-19}, headlines containing specific keywords like \texttt{covid}, \texttt{covid-19}, \texttt{global pandemic}, and \texttt{coronavirus} were tracked. 
These centrality measures appear to effectively capture significant moments in the pandemic timeline.

\subsection{FinDKG-based thematic investing} 
\label{sec:findkg_investing}

Thematic investing is an investment strategy that targets specific themes or trends that are anticipated to influence the future landscape of industries and economies. We demonstrate the utility of FinDKG and link prediction with KGTransformer in estimating corporate exposure to AI, %a theme that has seen heightened interest 
increasingly popular since the launch of OpenAI’s ChatGPT. The objective is to quantitatively measure how closely aligned stock entities are to the prevalence of the AI theme and to generate forward-looking exposure scores. 

In an online learning setting, we fit a KGTransformer model within the three-year rolling window FinDKGs at the end of every quarter. At each time $t$ (corresponding to the end of each month), the fitted KGTransformer is used to predict which stock entities are likely to be impacted by AI in the upcoming period $t+1$, corresponding to the quadruple (\texttt{AI}, \texttt{Impact}, ?, $t+1$). Only stocks with a predicted impact likelihood exceeding the average across all entities are retained. This selection forms the basis of a monthly-rebalanced, AI-focused long-only portfolio within the US S\&P 500. The portfolio is constructed by using the normalised predicted likelihood scores as the holding weight, which sum up to 100\%. We denote this KGTransformer-powered portfolio as FinDKG-AI.
We also fit an EvoKG model as an alternative KG learning methodology, using the same FinDKG data and settings, thereby constructing an EvoKG-based AI portfolio as a baseline strategy.

\begin{figure}[t]
  \centering
  \includegraphics[width=.95\linewidth]{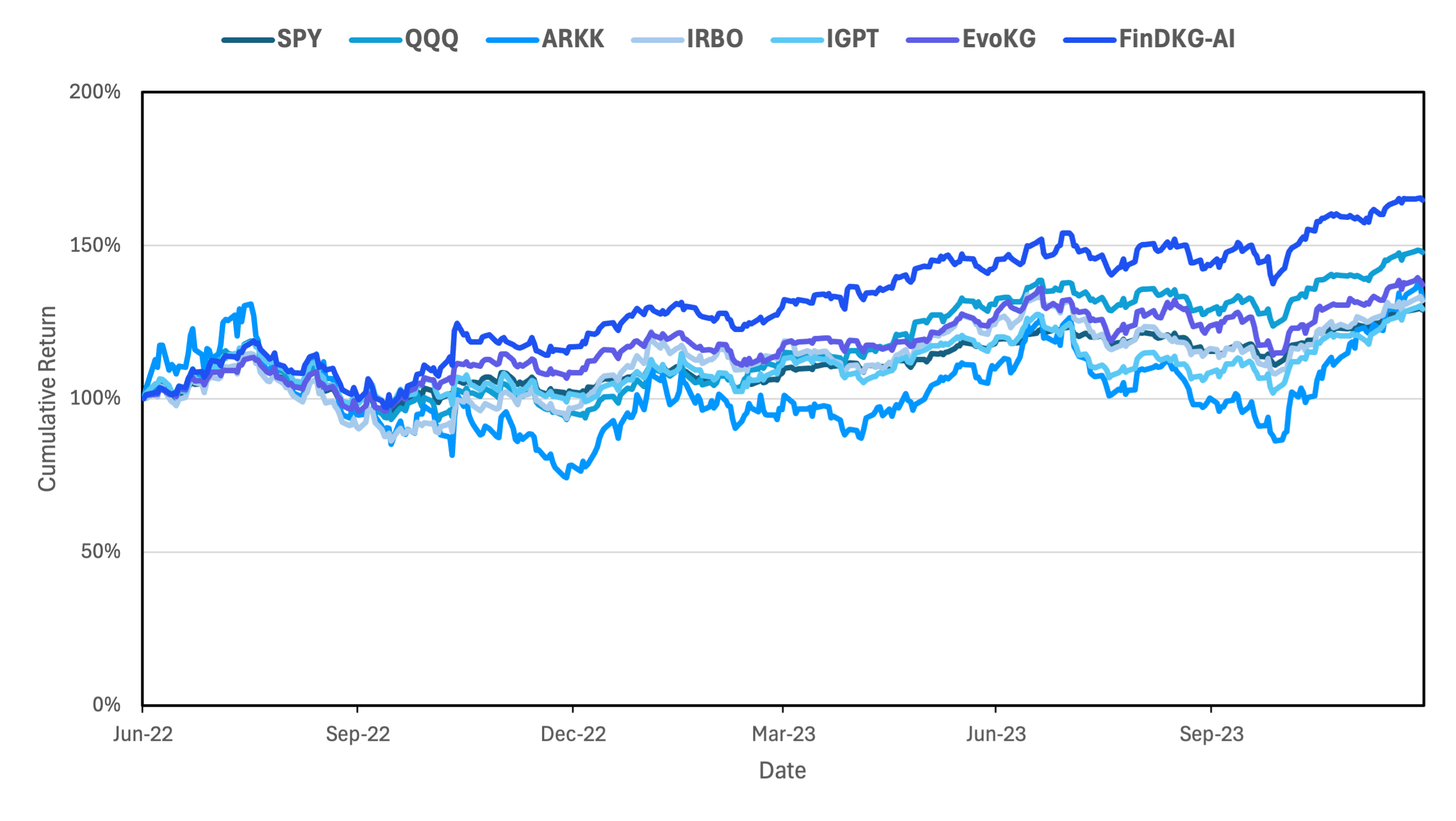}
  \caption{Cumulative returns of AI-themed long-only portfolios and market indices from June 2022 to December 2023.}
  \label{fig:wealth}
\end{figure}

The out-of-sample backtesting results in Table~\ref{tab:backtest} show the efficacy of the FinDKG-based AI portfolio: % for thematic investing: 
FinDKG-AI achieves the highest annualized return and Sharpe ratio across all portfolios. The existing AI ETFs lag behind the market benchmark with less return and comparably larger risk. In contrast, the KGTransformer-based FinDKG AI portfolio outperforms competitors %significantly %in the recent AI trend 
across the evaluation period, with a jump coinciding approximately with the release of OpenAI's ChatGPT in November 2022, as shown in Figure~\ref{fig:wealth}. The FinDKG-AI portfolio also outperforms the EvoKG-based strategy, aligning with the better link prediction capabilities of the KGTransformer model architecture shown in previous section.

\begin{table*}[t]
\centering
\scalebox{0.9}{
\begin{tabular}{lccccc}
\toprule
Portfolio & Name / Model & Annualized Return & Annualized Volatility & Sharpe Ratio & Max DD \\
\midrule
SPY & SPDR S\&P 500 ETF & 18.6\% & \underline{\textbf{17.1\%}} & 1.084 & \underline{\textbf{-16.7\%}} \\
QQQ & Invesco NASDAQ-100 ETF & \textbf{29.8\%} & 22.5\% & \textbf{1.323} & -21.6\% \\
ARKK & ARK Innovation ETF & 20.6\% & 47.9\% & 0.431 & -43.2\% \\
IRBO & Invesco AI and Next Gen Software ETF & 20.4\% & 25.9\% & 0.786 & -24.9\% \\
IGPT & iShares Robotics and AI ETF & 18.7\% & 22.8\% & 0.820 & -20.0\% \\ 
EvoKG & FinDKG EvoKG-predicted Portfolio & 23.5\% & \textbf{20.2} \% & 1.163 & -19.1\% \\
FinDKG-AI & FinDKG KGTransformer-predicted AI Portfolio & \underline{\textbf{39.6\%}} & 21.9\% & \underline{\textbf{1.810}} & \textbf{-18.2\%} \\
\bottomrule
\end{tabular}
}
\caption{Overall performance of market, AI-themed ETF, and FinDKG portfolios. The top two performing portfolios within the metric are highlighted in bold, and the best one is further underlined. The evaluation period is from 30/06/2022 to 29/12/2023.}
\label{tab:backtest}
\end{table*}

\section{Conclusion} %and discussion}
\label{sec:conclusion}

In this work, we provided three contributions around the use of dynamic knowledge graphs (DKGs) and large language models (LLMs) within financial applications. First, we investigated the performance of fine-tuned open-source LLMs in generating knowledge graphs, proposing the novel open-source Integrated Contextual Knowledge Graph Generator (ICKG) LLM. Next, the ICKG LLM is used to create an open-source dataset from a corpus of financial news articles, called FinDKG. 
Additionally, we proposed an attention-based architecture called KGTransformer, which incorporates information from meta-entities within the learning process, combining architectures such as HGT \cite{hu2020heterogeneousgraphtransformer} and EvoKG \cite{park2022evokg}. 

Our findings show that the proposed KGTransformer architecture improves the state-of-the-art link prediction performance on two benchmark datasets, and it achieves the best performance with over 10\% uplift on FinDKG. The generalizability of the ICKG LLM extends beyond the financial news and financial domain, as evidenced by applications in the recent literature adopting similar frameworks \cite{sarmah2024hybridrag,ouyang2024modal}. Code associated with this work can be found in the GitHub repository \ifthenelse{\boolean{anonym}}{ANONYMISED LINK}{\href{https://github.com/xiaohui-victor-li/FinDKG/tree/main}{xiaohui-victor-li/FinDKG}}, and an online portal to visualise FinDKG is available at \ifthenelse{\boolean{anonym}}{ANONYMISED LINK}{\href{https://xiaohui-victor-li.github.io/FinDKG/}{https://xiaohui-victor-li.github.io/FinDKG/}}.

\begin{acks}
FSP acknowledges funding from the EPSRC, grant no. EP/Y002113/1.
\end{acks}

\bibliographystyle{ACM-Reference-Format}
\bibliography{biblio}

%\appendix

\end{document}